\documentclass[aps,twocolumn,superscriptaddress,preprintnumbers]{revtex4}
\usepackage{epsfig}
\usepackage{graphicx,amssymb,epsfig,amsmath}

\newcommand{\be}{\begin{eqnarray}}
\newcommand{\ee}{\end{eqnarray}}
\newcommand{\ket}[1]{|#1\rangle}

\begin{document}

\title{Relation between semi- and fully-device-independent protocols}

\author{Hong-Wei Li$^{1,4,5}$, Piotr Mironowicz$^{2,6}$, Marcin Paw{\l}owski$^{3,2}$, Zhen-Qiang Yin$^{1}$, Yu-Chun
Wu$^1$, Shuang Wang$^{1}$, Wei Chen$^{1}$, Hong-Gang Hu$^4$,
Guang-Can Guo$^1$,  Zheng-Fu Han}

 \affiliation
 {Key Laboratory of Quantum Information,University of Science and Technology of China,Hefei, 230026,
 China\\
 $^2$ Instytut Fizyki Teoretycznej i Astrofizyki, Uniwersytet
Gda\'{n}ski, PL-80-952 Gda\'{n}sk, Poland
\\$^3$  Department of Mathematics, University of Bristol,
Bristol BS8 1TW,United Kingdom
\\$^4$ Department of Electronic
Engineering and Information Science, University of Science and
Technology of China, Hefei, 230027, China
\\$^5$ Zhengzhou Information Science and Technology Institute, Zhengzhou, 450004,
 China
\\ $^6$ Department of Microwave and Antenna Engineering, Faculty of Electronics, Telecommunications and Informatics, Gdansk University of Technology, Gdansk 80-233, Poland
 }

\begin{abstract}
We study the relation between semi and fully device independent protocols. As a tool, we use the correspondence between Bell inequalities and dimension witnesses. We present a method for converting the former into the latter and vice versa. This relation provides us with interesting results for both scenarios. First, we find new random number generation protocols with higher bit rates for both the semi and fully device independent cases. As a byproduct, we obtain whole new classes of Bell inequalities and dimension witnesses. Then, we show how optimization methods used in studies on Bell inequalities can be adopted for dimension witnesses.
\end{abstract}

\maketitle

{\it Introduction - } In device-independent (DI) protocols, two distant parties either do not know all the relevant parameters of their machines or do not trust them. This was formally presented in \cite{DI1}. Initially this approach was very
successful in quantum cryptography \cite{DIqkd1,DIqkd2,DIqkd3,DIqkd4}. Later, Colbeck \cite{Colbeck 1,Colbeck
2} proposed a true random number expansion protocol based on
the GHZ test, while Pironio {\it et al}. \cite{Pironio 1} proposed
a protocol based on Bell inequality violations. All these
protocols require entanglement, which has a negative effect on the
complexity of the devices and the rates of randomness generation \cite{Pironio 1} and key distribution.
To cope with this problem the semi device-independent (SDI) scenario was introduced in \cite{Witness}. In this approach, we consider prepare, and measure protocols without making any assumptions about the internal operations of the preparation and measurement devices. The only assumption made is about the size of the communicated system. We assume there to be a single qubit in each round of the experiment. This approach is a very good compromise between the fully DI scenario and experimental feasibility. The possibility of using prepare and measure protocols implies no need for entanglement, which makes the experiments easier by several orders of magnitude. However, the price to pay for this is that one extra assumption means the possibility of a loophole if not met. This lowers the overall security of the protocol, albeit not significantly, since it is relatively easy to find the dimension of the system in which Alice's device encodes information even through superficial inspection of the device. However, it is almost impossible to test each part of the device to check whether it indeed works as advertised.
The first SDI protocol, presented in \cite{Witness}, was for quantum key distribution. Shortly thereafter, the first SDI randomness expansion protocol was proposed \cite{Li1}.
This work studies the relation between DI and SDI protocols. We show how and under what conditions one can be converted into the other and how this change affects their parameters. This relation provides us with interesting results for both scenarios. First, we find new random number generation protocols with higher bit rates for both semi and fully device-independent cases. As a byproduct we obtain whole new classes of Bell inequalities and dimension witnesses. Then, we show how optimization methods used in studies on Bell inequalities can be adopted for dimension witnesses.
Our paper is structured as follows. First we describe the method for converting DI protocols to SDI and vice versa. Then we apply our method to SDI random generators to obtain new DI protocols with higher bit rates. We also present a new family of Bell inequalities. Next we take a class of DI protocols and turn these into SDI protocols with better rates. This time our byproduct is a new family of dimension witnesses. Finally, we show how semi definite programming (SDP) methods, which are a powerful tool in the DI scenario, can be used in an SDI one.

{\it Bell inequalities and dimension witnesses - } In a DI protocol, distant parties receive systems in an unknown (possibly) entangled
state from an untrusted sender. In each round they choose their
inputs and make measurements to obtain the outcomes. In our paper we
are interested in bi-partite protocols, and thus, we have two parties:
Alice and Bob, with their setting choice denoted by $x$ and $y$, respectively, and their outcome by $a$ and $b$, respectively. In some, randomly
chosen rounds of the protocol, both parties will publicly compare
their settings and outcomes to estimate the conditional probability
distribution $P(a,b|x,y)$. From this they can calculate the value of
some Bell inequality \be \label{bi}
I=\sum_{a,b,x,y}\alpha_{a,b,x,y}P(a,b|x,y), \ee
which is their security parameter.
This parameter can then be used as the lower bound on the amount of
randomness or secrecy in the remaining rounds.
In an SDI protocol Alice chooses her input $x'$, but she does not have
any outcome. Instead, in each round, she prepares a state depending
on $x'$ and sends it to Bob. Bob chooses his measurement setting $y$ and
obtains outcome $b$. Although the devices that prepare the system and
then measure it are not trusted, we assume that the communicated
states are described by a Hilbert space with a fixed dimension (here
we assume they are qubits) and that there is no entanglement between
the devices of Alice and Bob. Again in some rounds, $x', y$, and $b$
are announced to estimate the value of some dimension witness \be
\label{witness} W=\sum_{b,x',y}\beta_{b,x',y}P(b|x',y), \ee which
has exactly the same function as $I$ in the DI case. Both of these
scenarios are illustrated in Fig. 1.

\begin{figure}[!h]\center
\resizebox{9cm}{!}{
\includegraphics{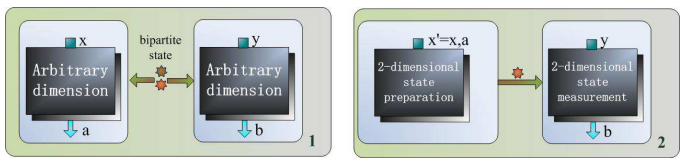}}
\caption{(Color Online) Schematic representation of DI (1) and SDI (2) protocols and of our method for finding the corresponding ones.}
\end{figure}

Dimension witnesses were introduced in \cite{Gallego}. Just as
violation of a Bell inequality in the DI case tells us that the measured
system cannot have a classical description, violation of a
dimension witness in the SDI case tells us that the communicated system
cannot be a classical bit (in the case of the witness for dimension
2). In both cases violation of the classical bound is a necessary
(though not always sufficient) condition for the protocol to work.
Moreover, in both cases the form of $I$ or $W$ is the most
important part of the protocol's description. Therefore, finding
the correspondence between these two objects is equivalent to finding
the correspondence between the protocols. Our method for doing so is
quite straightforward:
Let us rewrite $I$ as
$\sum_{a,b,x,y}\alpha_{a,b,x,y}P(a|x,y)P(b|a,x,y)$ and start by
considering $a$ as part of Alice's input. This is a purely
mathematical operation, and has no meaning at the protocol level.
Now Alice's input is $x'=(x,a)$. We can consider $P(a|x,y)$ as the
probability that part of Alice's input is $a$. Because in the
parameter estimation phase of the protocol the inputs are chosen
according to a uniform distribution, we set $P(a|x,y)=\frac{1}{A}$,
where $A$ is the size of the alphabet of $a$. Our $I$ is now
$\sum_{b,x',y}\alpha_{b,x',y}\frac{1}{A}P(b|x',y)$ and has the form
of (\ref{witness}) with $\beta_{b,x',y}=\frac{1}{A}\alpha_{b,x',y}$.
Our method is quite heuristic and there is no guarantee that a Bell
inequality with a quantum bound higher than the classical one will lead to a
dimension witness that can be violated. Also using it to go from
a dimension witness to a Bell inequality is not always possible. To do
so, Alice's input $x'$ must be divided into a pair comprising a setting and an
outcome. This is only possible if the alphabet of $x'$ has a composite
size. These are serious drawbacks, but they are easily outweighed by
the advantages: simplicity and the fact that the method works! In the following
paragraphs we apply it to generate new useful witnesses,
inequalities, and protocols.

{\it From SDI to DI protocols - }Let us consider the family of SDI protocols for randomness
generation introduced in \cite{Li2}, which are based on $n\to 1$ quantum
random access codes \cite{RAC2}. Alice's input $x'$ is a collection
of $n$ independent bits $a_0,...,a_{n-1}$. For Bob $y=0,..,n-1$. The
dimension witness is defined by $\beta_{b,x',y}=\delta_{a_y,b}$.
There are many ways of dividing Alice's input into pairs of
settings and outcomes but, because of the independence of the bits,
they are all equivalent. Let us then take outcome $a$ to be $a_0$ and
setting $x$ to be $a_1,...,a_{n-1}$. In this way we obtain a new
family of Bell inequalities
\be
I_n=\sum_{a,b,x,y}\delta_{a_y,b}P(a,b|y,x).
\ee
Systems obtaining
a high value of $I_n$ can be used to implement entanglement assisted
random access codes \cite{M-EARAC}. In these codes Alice has $n$
independent bits and Bob is interested in only one of them. Alice
can send only one bit of classical communication to Bob, but they can
share entanglement. If we denote the bits that Alice wants to encode
by $c_0,...,c_{n-1}$, then Alice can choose her setting by taking
$a_i=c_i\oplus c_0$ for all $i>0$ and transmit the message
$m=a\oplus c_0$ to Bob. If he XORs his outcome $b$ with the message
it is easy to calculate that he obtains the correct value of $a_y$ with average
probability $P_n=\frac{I_n}{n2^n}$. Therefore, we see that there is indeed a correspondence between the dimension witness and the Bell inequality related by our method, also at the level of protocols. In this case they are both a measure of the success probability for the different kinds of random access codes. $I_2$ is equivalent to the CHSH inequality. However, members of this family for $n>2$ have never been studied. Because it is possible to use them for entanglement assisted random access codes, the bounds on their efficiency derived in \cite{M-EARAC} apply and they translate to the maximum quantum value of $P_n$, that is, $P_n^{max}=\frac{1}{2}\left(1+\frac{1}{\sqrt{n}} \right)$.
Now we show how our new Bell inequalities perform in DI randomness generation. The quantity that we wish to optimize is the min-entropy $H_{\infty}(a,b|x,y)=-\log \max_{a,b} P(a,b|x,y)$. To find the lower bound on this for a given value of $P_n$ we use the methods described in \cite{NPA}. More precisely, we bound the set of allowed probability distributions by the second level of their hierarchy. We obtained the following lower bounds on the min-entropy for the maximal quantum values of $P_n$:

\begin{table}[h]
\begin{tabular}{c|c|c}
\hline
$n$ & DI: $H_{\infty}(a,b|x,y)$ & SDI: $H_{\infty}(b|a,x,y)$ \\
\hline
2 & 1.2284 & 0.2284\\
3 & 1.3421 & 0.3425\\
4 & 1.4126 & 0.1388\\
5 & 1.4652 & 0.1024
\end{tabular}
\caption{Lower bounds on the min-entropy for the protocols corresponding to the $n \to 1$ random access codes. The values in the rightmost column are for the family of protocols defined in \cite{Li2} and are taken from there. The values in the middle column correspond to the min-entropy of the outcomes in Bell inequalities $I_n$ for the maximal quantum values thereof. These were obtained using the SDP methods in \cite{NPA}. The inequalities $I_n$ were derived from the protocols in \cite{Li2} using the method depicted in Fig. 1.}
\end{table}
Compared with the randomness obtained from the SDI protocols, the main difference is that it grows with $n$ instead of reaching a maximum at $n=3$. In fact the upper bound is $H_{\infty}(a,b|x,y)=1-\log P_n^{max}=2-\log\left(1+\frac{1}{\sqrt{n}}\right)$, which approaches 2 as $n\to \infty$. We conjecture that this is reached for any $n$ but the second level of the SDP hierarchy form \cite{NPA} that we use for the lower bound, is sufficient only for $n=2$. Proving this conjecture is one of the open areas of research. The lower bounds as a function of $P_n$ are plotted in Fig. 2.
\begin{figure}[!h]\center
\resizebox{8cm}{!}{
\includegraphics{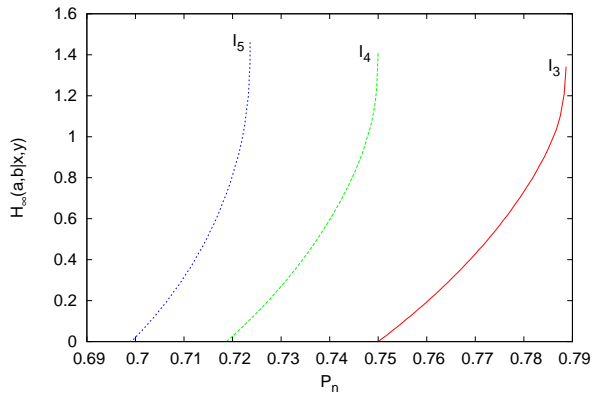}}
\caption{(Color Online) The lower bounds on $H_{\infty}(a,b|x,y)$ as functions of $P_n$.}
\end{figure}

{\it From DI to SDI protocols - } Now we apply our method to show that we can go the other way and convert a DI protocol to an SDI one. We start from the randomness generation protocol form \cite{twodi4} based on Bell inequality $I_{\alpha}$, which expressed in the form (\ref{bi}) is

\begin{equation}
\begin{array}{lll}
I_{\alpha}=\sum_{a,b,y}\delta_{a,b}\alpha P(a,b|x=0,y)\\~~~~~~~
+\sum_{a,b,y}\delta_{a,b\oplus y} P(a,b|x=1,y).
\end{array}
\end{equation}

Converting this to a dimension witness we get
\begin{equation}
\begin{array}{lll}
W_{\alpha}=\sum_{a,b,y}\frac{\alpha\delta_{a,b}}{2}
P(b|a,x=0,y)\\~~~~~~~~+\sum_{a,b,y}\frac{\delta_{a,b\oplus y}}{2}
P(b|a,x=1,y).
\end{array}
\end{equation}

The lower bound on the min-entropy as a function of coefficient
$\alpha$ is plotted in Fig. 3. For large values of $\alpha$ the
amount of randomness is clearly greater than that for the best of the
protocols described in \cite{Li2}.
The intuitive explanation for this is that $W_{\alpha}$ also corresponds to a kind of quantum random access code. In this case it is a $2 \to 1$ code with different weights assigned to the cases with $x=0$ or $x=1$. For large $\alpha$ it is much more important for the protocol to be correct when $x=0$ than in the case of $x=1$. This means that the protocols reaching maximum quantum value will tend to give the correct value of $b$ for $x=0$. Here correct means fully specified by $a, y$, and $x$, in other words, deterministic. The price paid for this is that for $x=1$ the probability of the correct (predetermined by $a, y$, and $x$) value is small, which implies a lot of randomness.
Previously in \cite{Li1,Li2}, the bounds on the entropy in SDI protocols were calculated using the
Levenberg$-$Marquardt algorithm \cite{LMA}, which is not guaranteed to
find global minima. SDP on the other hand always finds these; however, it was
previously not known how this could be applied in the SDI case. Below we give a
solution to this problem.

\begin{figure}[!h]\center
\resizebox{9cm}{!}{
\includegraphics{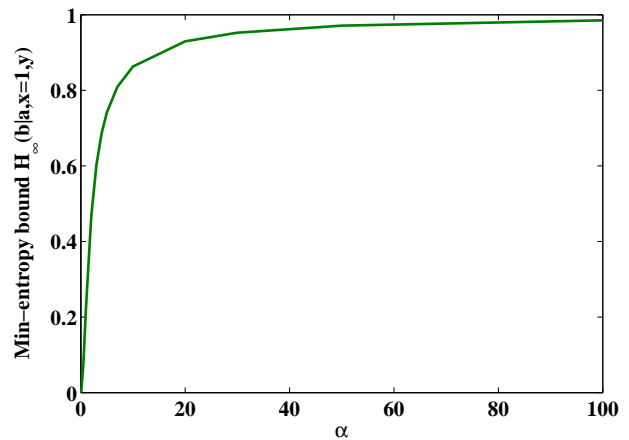}}
\caption{(Color Online) Lower bound on the min-entropy $H_{\infty}(b|a,x=1,y)$ as a function of
coefficient $\alpha$ for the maximal quantum value of $W_{\alpha}$. In \cite{twodi4} a large amount of randomness is generated only for one setting of $x$. Here we observe the same result with high randomness for $x=1$ and low randomness for $x=0$.}
\end{figure}

{\it Optimization in SDI protocols - }It is not possible to use SDP optimization directly in the SDI case because of the nonlinear target function. Neither can methods from \cite{NPA} be applied because they do not allow the dimension of the system to be set.
Therefore, we need to find another solution. We do it by proving the following theorem:

{\bf Theorem 1} {\it If $H_{\infty}(b|a,x,y)$ is the min-entropy obtained in the SDI case and $H_{\infty}(a,b|x,y)$ the min-entropy obtained in the corresponding DI protocol, then}
\be \label{had}
H_{\infty}(b|a,x,y)\geq H_{\infty}(a,b|x,y)-1,
\ee
{\it for the same value of the security parameter.}

{\it Proof - } See the appendix.

Let us stress that (\ref{had}) holds only when the values of the dimension witness and the Bell inequality are the same.
Consider Table 1 once again. For $n=2$ we have equality $H_{\infty}(b|x,a,y)=H_{\infty}(a,b|x,y)-1$. For $n=3$ $H_{\infty}(b|x,a,y)$ is slightly larger than $H_{\infty}(a,b|x,y)-1$. This most probably stems from the fact that the bound in the table is not tight for $n=3$. In fact, the upper bound on $H_{\infty}(a,b|x,y)$ is exactly $H_{\infty}(b|x,a,y)+1$. The situation changes for $n=4,5$. In these cases (\ref{had}) does not seem to hold. This is because the values in the table are given for the maximal quantum values of witnesses and inequalities which, for $n=4,5$ are not the same. If we calculate the entropy bound for the DI case when the value of the Bell inequality is equal to the maximal quantum value of the dimension witness, then the values are in agreement with (\ref{had}).
Using this method we were able to refine the results in \cite{Li1}, as shown in Fig. 4.

\begin{figure}[!h]\center
\resizebox{9cm}{!}{
\includegraphics{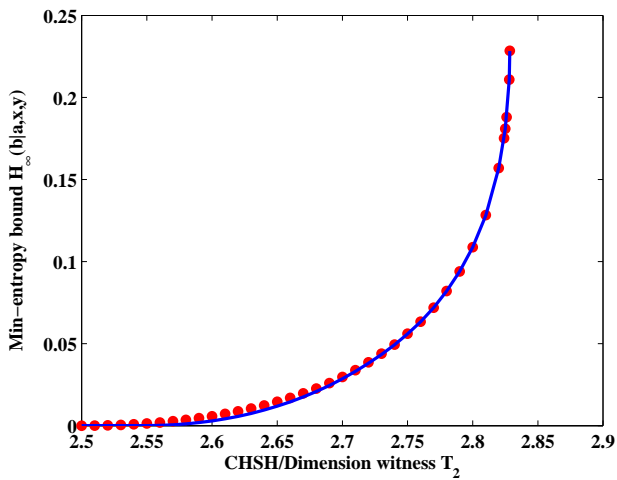}}
\caption{(Color Online) Min-entropy bounds for the SDI randomness generation protocol
based on the $2\to 1$ quantum random access code. The dots are obtained from
the Levenberg$-$Marquardt algorithm used in \cite{Li1}, which
is not guaranteed to find global minima, while the line depicts the SDP method described here. Note that state preparation
in the SDI protocol assumes that $p(a|x,y)=\frac{1}{2}$.}
\end{figure}

{\it Conclusions - } We investigated the relation between DI and SDI protocols. Although our study focused on randomness generation, our results are also applicable to quantum key distribution since all the state-of-the-art proofs of security are based on the randomness of measurement outcomes \cite{DIqkd2,DIqkd3}. To this end we demonstrated a method for converting Bell inequalities into dimension witnesses and vice versa. This allowed us to generate new examples of both types of objects with very interesting properties. Our new family of Bell inequalities gave rise to DI randomness generation protocols with better bit rates, while our family of dimension witnesses did the same for SDI protocols. Finally, using the correspondence between the DI and SDI approach we were able to modify the SDP-based methods, which were proven successful in the former case, to work in the latter one.
Apart from the similarities, our study also showed interesting differences such as the completely different dependence on $n$ in Table 1. It also introduced many new protocols for both scenarios. Comparison of their efficiency with that of existing ones, especially in the presence of noise and imperfect detectors, opened a new area of research.

{\it Acknowledgements -} H-W.L. wishes to thank YaoYao for his helpful discussion.
This work has been supported by the National Natural Science Foundation of China (Grant Nos. 61101137,
61201239, 61205118, 10974193, and 11275182), UK EPSRC, FNP TEAM and ERC grant QOLAPS. SDP was implemented in MATLAB using toolboxes \cite{SeDuMi,Yalmip}.

\section{Appendix: Proof of Theorem 1}

Every SDI protocol can be realized in the following way. Alice has a
pair of systems in the singlet state. If she wishes to prepare
state $\ket{\phi}$, she measures one particle in the basis
$\{\ket{\phi},\ket{\phi^{\perp}}\}$ and the other will also collapse to
one of these states. Based on her measurement outcome she either
sends the other particle to Bob unchanged or performs the unitary that
flips $\ket{\phi^{\perp}}$ to $\ket{\phi}$ and then sends it. If
Bob's measurement outcomes are binary Alice does not even have to perform
this unitary. She can just send her measurement outcome to
Bob (0 denoting $\ket{\phi}$ and 1 denoting $\ket{\phi^{\perp}}$)
who after XORing it with his outcome will get exactly the same
probability distribution $P(b|a,x,y)$ as in the initial SDI
protocol. These two cases are demarcated by letters (A) and (B) in Fig.
1.
\begin{figure}[!h]\center
\resizebox{8cm}{!}{
\includegraphics{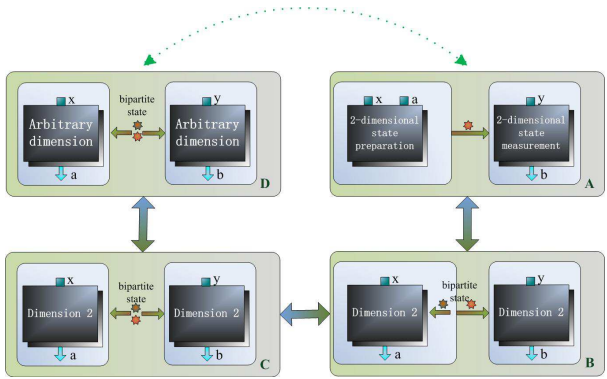}}
\caption{Overview of the method for lower bounding the entropy in SDI protocols using SDP. See the text for a detailed description.}
\end{figure}
Obviously, nothing changes if the source of the singlet states is
outside Alice's lab and she is only the receiver of one of the subsystems
just like Bob. This is depicted in fragment (C). But a lot changes
if we now assume that the state that they receive can be an arbitrary
maximally entangled state of any dimension (D). However, this only
enlarges the space of allowed probability distributions
$P(b|a,x,y)$, so any lower bound on the entropy in case (D) will
also hold in (A). Finally, (D) is just the description of a DI
protocol with some additional assumptions on the state. We can
lower bound the entropy in this case with the SDP-based methods in
\cite{NPA} and the fact that the state is maximally entangled will
be reflected by adding constraints
\be
\forall_{x,y} \quad P(a|x,y)=\frac{1}{2}.
\ee
We now have
\be \nonumber
H_{\infty}^{(A)}(b|x,a,y)\geq H_{\infty}^{(D)}(b|x,a,y)\\=-\log
\max_{b}P^{(D)}(b|x,a,y), \ee
which with
$P^{(D)}(b|x,a,y)=\frac{P^{(D)}(a,b|x,y)}{P^{(D)}(a|x,y)}=2P^{(D)}(a,b|x,y)$
gives \be \nonumber H_{\infty}^{(A)}(b|x,a,y)\geq -1-\log
\max_{b}P^{(D)}(a,b|x,y)
\\
\geq H_{\infty}^{(D)}(a,b|x,y)-1
\ee
The above formula implies that the randomness obtained in an SDI protocol is greater than or equal to that in its DI counterpart minus 1.

\end{document}